# Could Ball Lightning Be Magnetic Monopoles?

Author: Karl D. Stephan (Texas State University)

While magnetic monopoles have extensive theoretical justification for their existence, but have proved elusive to observe, ball lightning is both relatively frequently observed and largely unexplained theoretically. It was first proposed in 1990 that ball lightning might result from the catalysis of nuclear fission by a magnetic monopole. The observed frequency of ball lightning does not conflict with current upper theoretical or observational bounds for magnetic monopole flux. Some possible mechanisms to account for the association of magnetic-monopole-caused ball lightning with thunderstorms are described, and proposals for further observational and theoretical research are made.

## 1. Introduction

In this paper we examine the claim that the enigmatic phenomenon of ball lightning may be in fact a manifestation of effects caused by a magnetic monopole in free air. This claim connects a widespread and often-observed phenomenon—ball lightning—with a highly sought-after phenomenon—the magnetic monopole—which has been the subject of dozens of research projects in the last four decades. The current consensus regarding observational and experimental searches for magnetic monopoles is that no confirmed and repeatable observations have been made. The current consensus regarding ball lightning is that no theory covers most of the observed facts, but whatever mechanism produces ball lightning is probably of little fundamental physical significance and therefore does not justify much effort to explore. There is a small but non-zero probability that magnetic monopoles have been hiding in plain sight—as ball lightning.

Ever since Dirac's ingenious paper (Dirac, 1931) showed that at least one magnetic monopole in the universe would account for the quantization of electric charge, magnetic monopoles have been a theoretical possibility. But so far, attempts to observe magnetic monopoles have usually produced null results, which at most set an upper bound on the estimated flux of such particles, measured conventionally in $cm^{-2} sr^{-1} s^{-1}$. The well-known "Valentine's Day monopole" frustratingly detected exactly one time with Cabrera's superconducting-SQUID detector (Cabrera, 1982) was never replicated by him. Even when a different group of researchers in London with a similar setup detected one candidate monopole event a few years later (Caplin et al., 1986), the consensus now is that both of these events were possibly instrumental flukes rather than genuine magnetic monopoles. Reproducibility to the extent of detecting at least two similar events with the same experiment is a reasonable prerequisite for acceptance, but such reproducibility regarding magnetic monopoles has proved elusive up to now.

More recent searches for naturally occurring magnetic monopoles have concentrated on analyzing data from arrays of detectors set up primarily for other types of searches, such as the IceCube array at the South Pole (Lauber, 2021). There is also a possibility that collisions in the Large Hadron Collider (LHC) could create lighter varieties of magnetic monopoles (ATLAS Collaboration, 2023). These collider searches have also produced only negative results so far.

In contrast to magnetic monopoles, which are some of the most sought-after theoretical objects that have never been found, ball lightning is an empirical phenomenon that is encountered relatively often, but is still poorly understood theoretically. While a compilation of eyewitness accounts of ball lightning was published in English as early as 1855 (Arago, 1855), more recent books on ball lightning (e. g. (Barry, 1980; Boerner, 2019; Singer, 1971; Stenhoff, 1999)) show that in terms of the typical evolution of a scientific field, the study of ball lightning is still in its earliest stage, which consists in collecting and organizing data relevant to the phenomenon being investigated.

For those unfamiliar with ball lightning, a brief description follows. A typical ball-lightning sighting occurs when a thunderstorm is in the time and/or space vicinity of the eyewitness, who usually becomes aware of the object after it has formed. If the eyewitness is close enough to observe details, the object is typically perceived as a glowing spheroid having the total brightness of an incandescent light bulb with a consumed-power rating of 40 - 100 W. Reported colors tend toward the long-wavelength end of the spectrum (white, yellow, orange, red) with blue and green being less common. Typical diameters range from 20 - 1000 mm, with the peak of the reported distribution being around 160 mm ((Dijkhuis, 1992), cited in (Stenhoff, 1999)). While the object typically moves slowly (0.5 - 2 m s$^{-1}$) in mainly a horizontal direction, its brief lifetime (typically 1 - 15 s) and its typical altitude of less than 100 m means that remote-sensing technologies such as satellites and radar that have benefited other atmospheric-physics fields have not been useful for the investigation of ball lightning, although this state of affairs is beginning to change as security cameras and other video recording equipment become more widespread.

So far, most of what we know about ball lightning has been reported by untrained eyewitnesses who see it fortuitously. All attempts to reproduce it in the laboratory have failed to create an object which can show all of the above-mentioned characteristics. Although numerous experiments have produced objects having one or two characteristics of ball lightning, the laboratory objects typically last less than 1 s without the continuous application of external power, and either rise like the heated plasma which composes them (Egorov and Stepanov, 2002) or fall like burning liquid spheres of metallic material (Paiva et al., 2007). Our experimental studies (Stephan et al., 2013; Stephan and Massey, 2008) have explained several of these laboratory phenomena adequately in terms of known science, and show that any connection between them and the naturally-occurring phenomenon of ball lightning is tenuous at best. A recent review of research in this field (Shmatov and Stephan, 2019) shows that while there are literally dozens of ball-lightning theories, few of them account for more than one or two of the characteristic features of ball lightning, and none of them have led to the development of experiments which have successfully reproduced the phenomenon.

Several ball-lightning theories have received attention over the years, but have fallen out of favor as problems with them were pointed out. Nobelist Pyotr Kapitsa believed that the energy source for ball lightning could not be contained within its visible extent, and proposed that a naturally occurring radio-frequency energy source was responsible (Kapitsa, 1961). This theory fell into disfavor after improvements in RF receiving technology revealed that no such sources of high-power energy existed. Abrahamson and Dinniss proposed that conventional lightning strikes to certain types of soil would reduce $SiO_2$ to nanoparticles of elemental silicon, which would then burn to produce ball lightning (Abrahamson and Dinniss, 2000). Attempts to produce this phenomenon in the laboratory were unsuccessful, and although burning silicon has peculiar properties, it does not form ball lightning (Stephan and Massey, 2008). Numerous theories propose that ball lightning is essentially a plasma, but fail to explain adequately the source of energy needed to keep an atmospheric-pressure plasma excited to the point of light emission for several seconds (see e. g. (Dijkhuis, 1980; Wu, 2016)). Any chemical or plasma theory which requires that the same physical material remain inside the visible sphere as it travels cannot explain one particular well-attested property of ball lightning: its ability to pass through closed glass windows. The only physical entities which can easily pass through a few mm of solid glass are subatomic particles and electromagnetic or acoustic radiation. As we have recently shown (Stephan, 2024), the type of damage to windows occasionally caused by ball lightning implies that the entirety of the glass enclosed in the ball-lightning object is heated very rapidly throughout the glass volume. The ability of ball lightning to travel through several mm of glass cannot be explained by the majority of theories which have been proposed to explain ball lightning, which remains largely as enigmatic as it was at the dawn of the scientific revolution.

We will now present the cases both for and against the proposed connection between ball lightning and magnetic monopoles, beginning with evidence favoring the connection.

## 2. Evidence for identifying ball lightning with magnetic monopoles

### 2.1. Statistical evidence

We begin with what *is* known about magnetic monopoles, namely that if they exist at all, they are rare in the vicinity of Earth. A review of monopole research up to 2012 (Rajantie, 2012) describes both theoretical and observational upper bounds for the estimated magnetic monopole flux at the Earth's surface. While most searches assume that the magnetic monopole sought has only 1 to 3 units of the fundamental quantum of magnetic charge posed by Dirac (equal to 4.14 x $10^{-15}$ Wb in SI units), the range of masses that various theories predict is quite large. In energy terms, magnetic monopoles as light as 10 GeV or as heavy as $10^{16}$ GeV are predicted by various theories (Rajantie, 2012).

Direct experimental searches can be divided into searches for naturally occurring magnetic monopoles and searches for those produced artificially in particle accelerators. Because the highest energy accessible to current particle accelerators is ~13 TeV, only the lighter magnetic monopoles could be produced with them, and accelerator-based searches have thus far been unsuccessful.

Some searches for naturally occurring magnetic monopoles rely on the particles' assumed high energies, with $\beta = v/c$ exceeding 0.1, and rely on typical high-energy particle detection methods such as etching tracks in solids permeated by the particles or detecting scintillations from ionization caused by the particle. Other searches use the fact that a magnetic monopole's magnetic field will produce a persistent current in a superconducting loop through which the particle passes, and this current can then be detected with a superconducting quantum-interference detector (SQUID). This type of detector was used in (Cabrera, 1982) and (Caplin et al., 1986), and does not depend on the magnetic monopole's kinetic energy at all.

As summarized in (Rajantie, 2012), the flux of intermediate-mass magnetic monopoles on the Earth's surface has a probable upper bound on the order of $10^{-16}$ cm$^{-2}$ sr$^{-1}$ s$^{-1}$, while the upper bound for massive Grand-Unified-Theory (GUT) monopoles (the $10^{16}$-GeV type) is lower: about $10^{-19}$ cm$^{-2}$ sr$^{-1}$ s$^{-1}$. To put the latter flux in perspective, for the GUT upper bound a 100-cm$^2$ detector would have a chance of detecting a GUT magnetic monopole about once in every 300 million years. This is one reason why the early results of Cabrera and Caplin have been cast into doubt, but the field of monopole searches is far from mature and unknown factors may be at work which may eventually invalidate the current upper bounds.

If we take the best current estimates of upper bounds for magnetic monopole flux and compare them to the best current estimates of the frequency of occurrence of ball lightning, the ball-lightning-as-magnetic-monopole hypothesis will be in trouble if ball lightning turns out to be a lot more frequent than the upper bounds for magnetic monopoles. Fortunately, this is not the case.

Reliable estimates of the frequency of ball lightning are not easy to make. Nevertheless, we have recently found a method involving citizen-science efforts which can place a reasonably reliable *lower* bound on the frequency of ball lightning in the continental U. S.

Beginning in June of 2020, my colleagues Richard Sonnenfeld of New Mexico Tech, Alexander Keul of TU Vienna, and I have operated an Internet-based ball lightning report website (https://tinyurl.com/BLReport) at which individuals wishing to file a report of what they believe to be ball lightning can answer a series of questions and record a written narrative, in addition to providing objective data such as location, time, weather conditions, and other relevant circumstances. As of July of 2024, the site has logged over 900 separate entries. While assessing the quality of such reports will always involve a certain amount of subjectivity, we estimate that over 50% of the reports probably represent what in the present state of the science we can term genuine ball lightning. There are many phenomena that appear to be ball lightning to the untrained observer, but can be explained with more mundane causes such as aircraft lights, meteors, and power-line arcs triggered by lightning. Even when these cases are eliminated from our raw reports, this means that we have records of about 400 cases of ball lightning in the database.

Of the reports which pertain to cases that occur in a given time frame, if we grant the assumption that the cases are *bona fide* ball lightning, our database provides us with a means of setting an objective lower bound on the frequency of ball lightning in a given geographic area. We have carried out this procedure on a sample of reports that related to ball lightning in the continental U. S. that occurred between Jan. 1 and Aug. 31 of 2022. These reports were filtered by means of the following criteria:

1. The location is within the 48 contiguous U. S. states.
2. The location is specific to within a radius of approximately 1 km.
3. Only direct eyewitness reports are accepted (no third-party reports).
4. Only reports with time precision of ±2 months or better are accepted.
5. Only reports with occurrence dates between Jan. 1, 2022 and Aug. 31, 2022 are accepted.
6. Only reports for which the best explanation is ball lightning are accepted.

Of the 209 reports received in that 8-month time period (many of which dealt with historical events many years previous to the sample time), 37 reports met all six criteria listed above. Scaling this rate to a 12-month period yields an annual rate of 55.5 vetted incidents of ball lightning per year for the continental U. S. When one converts this figure to units conventionally used for magnetic-monopole flux, we come up with the surprisingly low figure of $1.75 \times 10^{-24}$ cm$^{-2}$ sr$^{-1}$ s$^{-1}$, or more than four orders of magnitude *lower* than the lowest upper bound for GUT monopoles of $10^{-19}$ cm$^{-2}$ sr$^{-1}$ s$^{-1}$.

The reason that no one has yet repeatably detected a magnetic monopole, yet we receive several reports of ball-lightning sightings every month even though their effective flux is much less than the upper bounds of magnetic-monopole flux, is that in a typical experimental search for naturally occurring magnetic monopoles, the detector is a single relatively small-area device.

But for ball lightning, we effectively have millions of detectors roaming the continental U. S. at no charge, on duty during their waking hours, and primed to notice and report the highly exotic and unusual phenomenon of ball lightning.  In a field where quantitative data is scarce, we nevertheless have a high degree of confidence in the lower bound of ball-lightning incidence calculated in this way.

The true frequency of ball lightning is undoubtedly higher than the lower bound we determined above.  Not everyone who sees ball lightning knows what it is and reports it to our website.  And of course, it is likely that many ball lightning objects appear in regions unobserved by anyone.  These two factors, while difficult to calculate, make it likely that the actual frequency of ball lightning is much greater than 55.5 per year in the continental U. S.

Approaching the problem from the other direction, if magnetic monopoles were visible, how frequently would they occur in terms of appearances per square kilometer per year, which is the standard metric for conventional lightning strikes?  The higher light-monopole upper bound of $10^{-16}$ $cm^{-2}$ $sr^{-1}$ $s^{-1}$ is equivalent to a flux of about 396 $km^{-2}$ $yr^{-1}$.  This is a factor of 10 greater than the average frequency of conventional lightning in the continental U. S., which is only about 24 $km^{-2}$ $yr^{-1}$.  As everyone agrees that ball lightning is much rarer than conventional lightning, this comparison shows that the upper bound for light magnetic monopole flux is much higher than the observed frequency of ball lightning by any reasonable estimate.

The supposition that some fraction of actual magnetic monopoles manifest themselves as ball lightning thus does not contradict any existing theoretical or experimental upper bound on the flux of magnetic monopoles.  The "flux" of ball lightning calculated from our database is lower than even the current lowest (GUT) upper bound of $10^{-19}$ $cm^{-2}$ $sr^{-1}$ $s^{-1}$ by several orders of magnitude.

A visual summary of these conclusions is shown in Fig. 1.  Even if we suppose that only 0.1% of the total of ball lightning occurrences in the U. S. were reported to our database in the sample period, the estimated frequency of ball lightning lies a factor of 10 below the lowest upper bound estimated flux of extremely high-energy magnetic monopoles, and 10,000 times below the upper flux bound of slower magnetic monopoles.

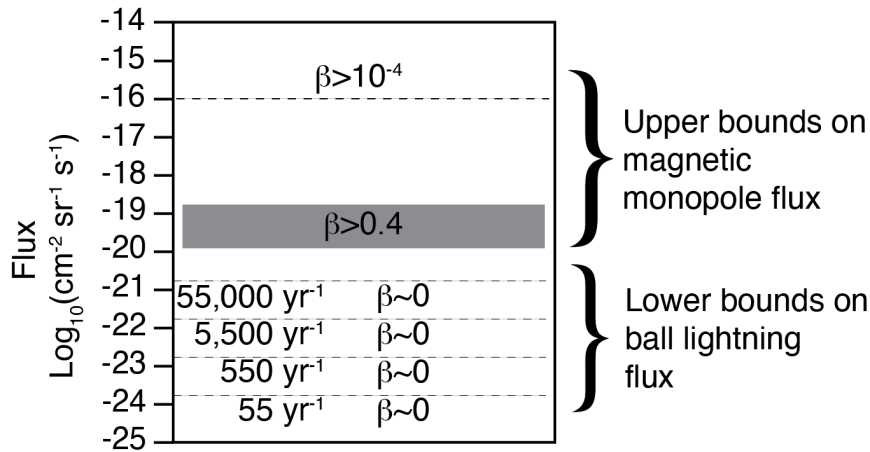

Fig. 1.  Upper bounds for flux of relativistic ($\beta > 0.4$) and slower ($\beta > 10^{-4}$) magnetic monopoles, (based on experiments described in (Rajantie, 2012) and (Hogan et al., 2008), respectively), and lower bounds for ball lightning flux.  The ball lightning flux is calculated from the stated number of occurrences per year in the continental U. S.

While this statistical compatibility is only a necessary condition for ball lightning to be associated with magnetic monopoles, it is one that is based largely on objective data and eliminates one possible objection to the hypothesis at the outset.

## 2.2. How magnetic monopoles could produce ball lightning

Next, we examine the reasoning behind the idea that a free magnetic monopole at rest in atmospheric-pressure air would produce visible phenomena resembling ball lightning. As far as we can determine, the first person to propose this hypothesis was Korshunov (Korshunov, 1990). Although the potential for magnetic monopoles to catalyze nuclear decay was recognized earlier (see e. g. (Lipkin, 1983)), Korshunov was the first to realize that this process might account for ball lightning. He examined what would happen to a GUT monopole, having a mass ~$10^{16}$ GeV and two Dirac-quantum magnetic charges, if it drifted in the atmosphere under the influence of the earth's magnetic field. He concluded that if any atomic nitrogen (N) were in the vicinity, the nucleon's significant magnetic dipole moment would cause a concentration of such atoms around the magnetic monopole. He estimated that the magnetic monopole would initiate a series of nuclear reactions in nearby N nuclei that would terminate in decomposition into protons, neutrons, and other transient subatomic particles, and also would release a total of several hundred MeV for every N atom thus fissioned. At the rate he estimated these reactions would take place, a single magnetic monopole might release as much as 90 kW of energy, most of it contained in the kinetic energy of subatomic particles. He estimated that protons leaving the vicinity of the magnetic monopole would have an average energy of 3 MeV, and would thus penetrate atmospheric-pressure air to a distance of about 140 mm, which is well within the typical range of ball lightning radii.

The overall picture is of a magnetic monopole catalyzing what amounts to nuclear fission of atomic nitrogen, at a rate which would at least account for the optical power emitted by ball lightning. We should note that if the only power required to be accounted for is optical power, then a power level of as little as 4 W is adequate to account for most ball-lightning sightings, because a 60-W incandescent lamp emits only about 4 W in the optical-wavelength range, depending on design. So even if Korshunov's estimate of 90 kW is high by a factor of $10^4$, the basic mechanism could still account for the light emission of ball lightning.

The hypothesis that magnetic monopoles drive the phenomena of ball lightning has other explanatory powers besides accounting for light emission. One of the most enigmatic aspects of ball lightning is the fact that it appears inside structures without warning or preliminary effects, and without any apparent connection to an exterior energy source. Several well-known cases of ball lightning have occurred inside metal-skin aircraft, the most well-known being witnessed by physicist and radioastronomer R. C. Jennison (Jennison, 1969). We have numerous reports in our ball-lightning report database which describe ball lightning appearing inside houses and other structures.

While Korshunov did not address the question of whether a magnetic monopole can produce the above-noted fission effects while in rapid motion, there are reasons to believe it could not. The concentration in ambient atmospheric-pressure air of atomic nitrogen, as opposed to the dimer $N_2$, is vanishingly small (according to Boltzmann statistics, less that 1 atom per cubic km), although naturally occurring radioactivity produces a few N atoms per $cm^3$ by dissociating $N_2$ molecules. Once a magnetic monopole in air begins to catalyze a large number of N atoms per second, the ionization resulting from high-energy protons and other particles leaving its vicinity will produce enough atomic N by dissociation to keep the reaction going. But if the magnetic monopole is moving at a significant velocity ($> \sim 10$ m s$^{-1}$) with respect to ambient air that does not have a significant concentration of N atoms to begin with, the buildup of N atoms cannot take place and so the magnetic monopole can travel a large distance invisibly. Only when it (a) decelerates to $\sim < 10$ m s$^{-1}$ (a figure based on the observed velocity of ball lightning) and (b) encounters a significant concentration of N atoms, can the regenerative catalysis of N atoms and the accompanying visible air glow from ionization take place.

If we assume that a magnetic monopole is decelerating down to approximately zero net velocity (although Brownian motion appropriate to the temperature and particle mass will always be present), it will remain invisible while it passes through the atmosphere and is decelerated by solid objects such as walls, construction materials, and glass. Once it loses enough forward motion and encounters enough atomic N to start the regenerative fission process, it will become visible as ball lightning.

## 2.3. How ball lightning produced by magnetic monopoles can penetrate glass

As with other theories of ball lightning, passage through solid glass presents a challenge for the magnetic-monopole theory as well. But if we assume that a single magnetic monopole is the source of the energy that produces optical radiation, the power density near the core of the structure rises to high values. For example, if the total power emerging from the immediate vicinity of the magnetic monopole is as little as 1 kW, at a radius of 100 $\mu$m from the magnetic monopole the average power density is 7.9 GW m$^{-2}$. Such a concentration of energy is sufficient to vaporize any known material, and as long as the magnetic monopole itself has sufficient forward momentum, the power emitted from its vicinity could create a vapor-filled hole or tunnel in the glass. In a paper describing the careful examination of a piece of window glass through which ball lightning was observed to pass, Bychkov et al. (Bychkov et al., 2016) discovered a cylindrical hole 240 $\mu$m in diameter which did not completely penetrate the glass. Such a hole is consistent with the passage of a small but intensely powerful object, and it is possible that the molten glass around the hole partially closed it before solidifying. Radiation preceding the emergence of the magnetic monopole through the glass could provide enough atomic N to restart the fission reaction once the magnetic monopole emerges into clear air.

Sometimes ball lightning is observed to explode at the end of its life, with results that create damage comparable to that of a conventional lightning stroke. The encounter of a magnetic monopole with a solid substance whose elements are especially susceptible to being fissioned by magnetic monopoles could result in a rapid increase in the fission rate resembling an explosion. Alternatively, if a GUT type ($10^{16}$ GeV) north-pole-charged magnetic monopole encounters a south-pole-charged magnetic monopole which is its antiparticle, they will annihilate each other and release approximately 3 MJ of energy, which is comparable to the energy content estimated in some ball lightning incidents (Bychkov et al., 2002).

## 2.4. How the lifetime of ball lightning is explained with the magnetic-monopole hypothesis

Another characteristic of ball lightning is its typically brief lifetime. Rarely, a long-lived ball lightning object lasting 1 m or more has been reported, but the typical ball-lightning object is in view by the observer for only a few seconds before it either vanishes or explodes.

Theory predicts that magnetic monopoles are absolutely stable unless two oppositely-charged magnetic monopoles collide and annihilate each other. However, there are many ways a magnetic monopole could come to rest in matter without drawing attention to itself in the form of nuclear reactions or other energy-releasing processes. Because of the intense magnetic fields in the vicinity of a magnetic monopole, it will set up an image magnetic charge of opposite polarity in the surface of any ferromagnetic material, and will be attracted to it until the magnetic monopole crashes into the surface and becomes trapped at a location of locally high magnetic fields. In such a trapped state, it becomes what Rodionov terms "latent" (Rodionov, 1996) as opposed to "active" magnetic monopoles which are catalyzing nuclear decay. Another way a magnetic monopole can become trapped and latent is by entering a cavity inside a diamagnetic material. As most non-metallic substances are slightly diamagnetic, this mode of trapping is quite likely for a very-low-velocity magnetic monopole that encounters solid matter.

As we will see in the next section, the hypothesis that there are magnetic monopoles trapped on the surface of the earth in a latent state assists the explanation of why ball lightning is associated with thunderstorms.

## 2.5. Why magnetic-monopole ball lightning is associated with thunderstorms

As yet, we have not addressed the well-accepted observation that ball lightning tends to occur in the vicinity of thunderstorms. Thunderstorms have no obvious connection with magnetic monopoles, but two possible mechanisms for such an association can be supposed.

The first mechanism is the fact that thunderstorms produce relatively large volumes in which unusually high electric fields exceeding 5 kV m$^{-1}$ form and persist for minutes to hours at a time. As magnetic fields bend the paths of moving electric charges, electric fields can bend the paths of moving magnetic charges. While heavy (GUT) and high-energy magnetic monopoles are not significantly deviated by such fields, lower-mass and low-speed magnetic monopoles may be, and could possibly form the type of spiral track that is observed frequently in particle detectors when electrically charged particles enter a magnetic field. In following such a track, a magnetic monopole could lose most of its energy and slow down enough to become active and visible as ball lightning.

The second mechanism relates to the high peak currents that conventional lightning produces at and near the point on the ground where cloud-to-ground (CG) lightning strikes terminate. As measured by lightning-location-network instrumentation, many CG strokes produce instantaneous currents of 100 kA or more. Taking the typical diameter of a lightning channel to be 4 cm (Rakov and Uman, 2003), the maximum magnetic field at the edge of the channel is 1 T. With a typical return-stroke duration of 30 $\mu s$, such a field encountering a magnetic monopole already lodged in soil could impart energy to it ranging from a few keV to many GeV, depending on the monopole's mass. This assumes that magnetic monopoles are present in soil in some concentration. In a study of 23 kg of rock samples from polar areas of Earth, it was concluded that the absence of magnetic monopoles in the samples sets an upper concentration bound of 9.8 x 10$^{-5}$ g$^{-1}$ (Bendtz et al., 2013).

This is not a very restrictive upper bound when contrasted to the mass of earth within 100 mm of the surface over a 1 km$^2$ area. Assuming a conservatively low average mass density of 3000 kg m$^{-3}$, the mass of earth to a depth of 100 mm in 1 km$^2$ is 3 x 10$^{11}$ g. Assuming the upper bound measured by Bendtz et al. would predict a total number of 29.4 x 10$^6$ monopoles km$^{-2}$.

If we ask what the magnetic monopole concentration would have to be in order to produce, say, 55,000 ball lightning objects per year in the continental U. S. from CG lightning strikes, we can assume that the active volume within which the magnetic field from the strike could activate a magnetic monopole is a hemisphere 80 mm in diameter, with a volume of $1.07 \times 10^{-3}$ $m^{-3}$. The continental U. S. receives a minimum of about 25 million CG strikes per year (Vagasky et al., 2024), and if lightning does not strike the same place twice (and always hits the ground, which is not strictly correct), the total mass sampled by these strikes is therefore $8.04 \times 10^{10}$ g. Attributing 55,000 ball lightning events to magnetic monopoles implies an average concentration in soil of only $55,000/8.04 \times 10^{10}$ g $= 6.84 \times 10^{-11}$ $g^{-1}$, or more than six orders of magnitude lower than the upper bound established by Bendtz et al.

From the viewpoint of ball lightning theory, the need for a magnetic monopole to be present solves one of the persistent problems of ball lightning: while ball lightning is generally associated with thunderstorms, its frequency is much less than the frequency of conventional lightning, which implies that there must be a second ingredient present besides conventional lightning for ball lightning to form.

The hypothesis of magnetic monopoles is strengthened by incidents in which numerous ball lightning objects appeared apparently simultaneously, as in the cluster of sightings in Neuruppin, Germany in 1994 (Bäcker, 2007). The conditions at Neuruppin favored the propagation of "failed leaders"—lightning leaders emerging from the ground which failed to connect to cloud-based leaders, but which nevertheless could have carried significant currents at multiple locations (Stephan, 2023). If these currents produced magnetic fields of sufficient duration and intensity to dislodge multiple latent magnetic monopoles at relatively slow velocities, this would explain the near-simultaneous appearance of up to a dozen or so ball lightning objects in and around the town, some inside houses.

## 2.6. Magnetic monopoles powering ball lightning account for indications of ionizing radiation

Although it is not an effect which typically occurs during a sighting of ball lightning, there are a few well-attested incidents which provide evidence that ball lightning may emit ionizing radiation beyond the usual short-wavelength UV radiation to be expected from an atmospheric-pressure plasma. There is observational evidence (Stephan, 2024; Stephan et al., 2016) that ball lightning emits ionizing radiation: the incidents of window damage mentioned above, and the fact that ball lightning has been observed to cause fluorescence of a different color than the object itself. Shmatov (Shmatov, 2006) summarizes several incidents that imply ball lightning produces energetic ionizing radiation consisting of either high-energy charged particles or gamma rays. Several eyewitnesses have suffered illness resembling radiation sickness after being in close proximity to ball lightning. Two of the most significant incidents mentioned by Shmatov will be summarized here.

In 1965, physicist M. T. Dmitriev was camping near a river and brought along a number of instruments, including air-sampling equipment and a gamma-ray scintillometer. He witnessed the appearance of ball lightning over the river, and as it came ashore he managed to note the scintillometer reading and take air samples, which proved to contain a significant amount of ozone (Dmitriev, 1969, 1967). The exact reading of the scintillometer is uncertain, but it did indicate a burst of radioactivity in the vicinity.

And in 1886, a family in a hut outside Maracaibo, Venezuela saw a bright light and a "humming noise" for a short time, and then developed symptoms of what appeared to be acute radiation poisoning (Cowgill, 1886). As reported by a staff member of the U. S. consulate in that city who visited the nine people involved in the hospital, the members of the family suffered swellings, hair loss, and raw sores. There were no artificial sources of radiation in 1886, so the best explanation is that some naturally-occurring event caused the light, the sound, and the injuries. Ionizing radiation from an unusually intense ball lightning object is at least a possibility in this case.

If ball lightning is indeed a miniature fission reactor, why doesn't everyone who sees ball lightning in close proximity die of radiation poisoning? Considering the range of heavy charged particles in air, no one knows precisely the division of energy among gamma rays, neutrons, protons, electrons, positrons, and other more short-lived subatomic particles when a magnetic monopole catalyzes the fission of a nucleon. If we assume that the bulk of the energy is conveyed by charged particles the size of a proton or larger, the range of such particles in air is fairly short, even for energies of several MeV. According to the NIST PSTAR range tables for protons (https://www.nist.gov/pml/stopping-power-range-tables-electrons-protons-and-helium-ions), the projected range of a 3-MeV proton in dry sea-level air is 144 mm, meaning that anyone farther than that from the center of a ball-lightning object would receive essentially no charged-particle radiation. The projected range for 3-MeV alpha particles (helium nuclei) is only 17.3 mm, and proportionally less for larger nuclear fragments. While gamma rays and neutrons would be more penetrating, we can estimate the maximum possible exposure in grays (1 Gy = 1 J $kg^{-1}$) as follows.

Supposing the eyewitness to be 2 m away from the center of the ball-lightning object, let us assume that the power emitted during a 5-s lifetime is 1 kW. The approximate solid angle occupied by an adult human body at a distance of 2 m is (0.75 $m^2$/2 $m^2$) = 0.14 sr = $\Omega$. The fraction of isotropic power (of whatever form) received by that body is therefore ($\Omega$/4$\pi$) x 1 kW = 15 W. An exposure time of 5 s yields a maximum value for absorbed energy of 15 W x 5 s = 75 J. Assuming *all* the energy is in the form of radiation absorbed by a 68-kg body, the absorbed dose in grays is 1.1 Gy. A whole-body dose of 1 Gy, although serious and unsafe, has only about a 5% mortality rate. As the actual absorbed dose of ionizing radiation at this distance could be much less, it is likely that the majority of even close-range eyewitnesses would notice no immediate or long-term ill effects from radioactive ball lightning, although some would. And anecdotal evidence shows that some persons have suffered post-sighting ill effects that are consistent with mild to moderate radiation exposure (Shmatov, 2006). If the total power emitted from the object was closer to the maximum of 90 kW posed by Korshunov (Korshunov, 1990), radiation injury would be more likely. As most eyewitnesses sense no heat from ball lightning and only moderate light levels, it is more likely that the typical total power emitted from ball lightning is in the 10 W - 1 kW range, which mitigates the radiation hazard considerably.

3. Evidence against identifying ball lightning as magnetic monopoles

Here we consider the evidence opposing our hypothesis that ball lightning is produced by thermal-velocity magnetic monopoles in free air. We will discuss the relevant issues in the same order in which they were taken up in Section 2.

3.1. The proven scarcity of magnetic monopoles

Despite over fifty years of experimental and observational efforts, no one has yet produced convincing evidence that even two magnetic monopoles have been detected. The ball-lightning hypothesis requires magnetic monopoles to be present whenever ball lightning appears. And the incidence of ball lightning is high enough so that some estimates may approach or exceed the upper bounds already established for the flux of magnetic monopoles on Earth.

One way to estimate an upper bound of ball lightning frequency is to begin with an observation by Rakov and Uman (Rakov and Uman, 2003) that "from a few percent to about 10 percent" of individuals who live in regions with appreciable thunderstorm activity have seen ball lightning. As some fraction of U. S. residents live in areas where thunderstorms are rare, a reasonable estimate of the total U. S. population who have seen ball lightning is 1%. As the current U. S. population is about 330 million, this estimate implies that 3.3 million current U. S. residents have seen ball lightning at least once in their lifetimes. The average age of U. S. residents is about 38, and if the youngest age a person could recognize ball lightning (with help) is about 8, that leaves the average person about 30 years to observe ball lightning. This further assumption leads to an average annual rate of eyewitness ball-lightning incidents of 110,000 per year. If each incident is witnessed by two people on average, the actual number of incidents is cut in half, or 55,000 per year. Depending on what fraction of the continental U. S. is under observation at any given time, the true frequency of ball lightning could be as much as 10 to 100 times larger than this.

As there is only a factor of about $10^5$ separating the lower-bound ball lightning estimate of 55.5 per year in the continental U. S. from the lowest-bound estimate of the frequency of GUT-type magnetic monopoles, the true (but unknown) frequency of ball lightning may well exceed the lower-bound GUT magnetic-monopole flux estimate of $10^{-19}$ cm$^{-2}$ sr$^{-1}$ s$^{-1}$. If it does, then either the flux estimate is in error, or ball lightning has nothing to do with those types of magnetic monopoles. Unlike the question of magnetic monopole flux, the question of ball-lightning frequency can be addressed via more extensive and systematic surveys than we have heretofore performed. Although surveys can be costly, the resources required are miniscule compared to the cost of an LHC experiment, for example. If such surveys reveal that a new lower bound for ball-lightning occurrences exceed even the highest (light) magnetic-monopole flux estimate of $10^{-16}$ cm$^{-2}$ sr$^{-1}$ s$^{-1}$, then the connection between ball lightning and magnetic monopoles could be regarded as definitively severed, but not before.

## 3.2. Why magnetic monopoles probably do not produce ball lightning

While it is generally accepted that under the right conditions, magnetic monopoles could catalyze nuclear decay (or at least baryon decay), some of the theoretical upper bounds on occurrence of magnetic monopoles are based on the fact that monopoles are not destroying neutron stars and other vulnerable objects to any notable extent, or catalyzing the rapid heat destruction of the Earth.

If we consider only the magnetic monopoles that were created shortly after the Big Bang, they would be accelerated by interstellar magnetic fields to velocities close to the speed of light. Especially if they are the GUT (heavy) variety, such high-velocity magnetic monopoles are capable of penetrating the entire Earth without losing significant energy. If most magnetic monopoles in our vicinity have velocities exceeding about $\beta = 0.03$, they will behave like other high-energy cosmic-ray products and will never give rise to any phenomena that are directly visible with the naked eye, as ball lightning is.

Only lighter magnetic monopoles with fairly low initial velocities could be slowed down by ionization and elastic scattering to the extent of becoming thermal magnetic monopoles with essentially zero net velocity. And most researchers seem to assume that a thermalized magnetic monopole in liquid or solid matter will be trapped in the latent state without attracting attention by catalyzing nuclear fission. Even Korshunov (Korshunov, 1990), who first proposed the magnetic-monopole-ball-lightning connection, believes that a significant concentration of N atoms is necessary for an airborne thermalized magnetic monopole to begin catalyzing nuclear fission. The violent motion occasioned by even one such catalysis might disrupt the steady-state concentration of N atoms around the magnetic monopole to the extent that the reaction might cease before its energy production becomes significant or visible.

In short, the path to extensive ongoing catalysis of nuclear fission by a magnetic monopole is not well-analyzed theoretically, and claiming that it can happen in ambient air with only the presence of atomic N may be insufficiently justified by theory.

### 3.3. Why magnetic-monopole ball lightning would not penetrate glass

In the few cases where ball lightning has been witnessed to penetrate a closed glass window, its speed was slow enough to be tracked by the naked eye, which implies a velocity of <5 m s$^{-1}$ or so. The range of such a slow particle, even of the GUT variety, in a solid material such as glass is on the order of microns, not millimeters, which is the typical thickness of window glass. Even if we grant that a magnetic monopole could catalyze air molecules by dissociation and fission initially, the object's encounter with a solid surface such as glass is likely to quench the reaction and cause the object to lodge in the glass and become a latent magnetic monopole.

### 3.4. Why the lifetime of ball lightning argues against the magnetic-monopole hypothesis

As we noted above, magnetic monopoles are expected to be stable. If such an object managed to initiate self-perpetuating nuclear fission in air, there is no obvious reason why it would typically cease to do so after a time ranging from 5 to 60 s. Although some ball-lightning observations are terminated not by the end of the object's existence but by its passage out of the observer's line of sight, many such objects are observed by the eyewitness simply to vanish. A theoretical explanation of why a magnetic monopole can catalyze nuclear fission in air for only a few seconds is required, but nobody has proposed one as yet.

### 3.5. Why magnetic-monopole ball lightning should not be associated with thunderstorms

It seems that if ball lightning does occur when a magnetic monopole slows to a very low velocity in air, ball lightning should be a much more common occurrence than it is, and should not have a correlation with thunderstorms. If we consider magnetic monopoles arriving at the Earth's surface directly from space, they should be slowed by air and other material obstructions regardless of whether thunderstorms are nearby. Calculations indicate that the bending of a magnetic monopole's path by a typical thunderstorm electric field of 5 kV m$^{-1}$ results in a bending radius of 1 km or less if the velocity of the lightest type (~10-GeV) is only $10^4$ m s$^{-1}$. Such slow particles would be stopped by elastic collisions even before making one complete circle, so the electric-field-trap picture of why thunderstorms can trap magnetic monopoles appears to require some special conditions which may not apply to actual monopoles.

As to the hypothesis that the intense pulsed magnetic fields caused by CG lightning excite latent magnetic monopoles to become active, the monopoles first have to be present to be excited. The volume over which significant magnetic fields from CG lightning strikes occur is quite small, and it is not clear that the momentum imparted to a magnetic monopole in such a situation would not be dissipated in multiple collisions as heat, rather than resulting in a significant unidirectional momentum imparted to the particle that ejects it into the air.

3.6. Why ball lightning is probably not radioactive

Although there have been isolated incidents in which something resembling radiation poisoning has happened to eyewitnesses of ball lightning, we should bear in mind that closeup exposure to conventional lightning can mimic the burns, nausea, and some of the other ill effects caused by true ionizing-radiation poisoning. It is undoubtedly true that some conventional-lightning encounters were ascribed to ball lightning due to faulty recollection, viewing lightning leaders end-on, and other factors. If enough nuclear fission occurs to produce visible light with a brightness on the order of an incandescent lamp, it is likely that the accompanying ionizing radiation is more than enough to kill anyone within a meter or less of the object doing the radiating. Two criticality accidents occurred during and after the Manhattan Project during and after World War II, both due to accidents that pushed a plutonium nuclear bomb core into criticality (Anonymous, n.d.). One victim, H. K. Daghlian Jr. received an estimated 3.1 Gy of combined neutron and gamma radiation, and died 25 days later of acute radiation poisoning. The second victim, Louis J. Slotin, received about 10 Gy and survived only nine days. Both individuals were less than 1 m away from the core when it went critical.

There are numerous cases in the literature of ball lightning in which persons have come within less than 1 m from the object for times on the order of seconds. A young boy was directly struck by ball lightning, which left a Lichtenberg-figure pattern on his torso, but he apparently suffered no other ill effects (Garcia et al., 2018). If all *bona fide* ball lightning was powered by nuclear reactions, it seems unlikely that a person could have such a close encounter with it even momentarily without suffering a severe case of radiation poisoning.

4. Conclusions

We have considered a possible theoretical connection between a highly-sought and heretofore-undiscovered particle—the magnetic monopole—and a relatively-frequently observed phenomenon which has so far attracted little serious research effort—ball lightning. Because of the theoretical immaturity of ball lightning studies and the practical fact that no one has unequivocally observed or captured a magnetic monopole, the connection is not as firm as we would like. Nevertheless, there are theoretical arguments put forward by at least two researchers (Korshunov, 1990; Rodionov, 1996) which say that a free magnetic monopole nearly at rest in air could produce an energetic phenomenon resembling ball lightning.

In this paper, we have considered bodies of evidence which both favor the hypothesis of magnetic monopoles powering ball lightning by means of nuclear fission, and evidence which disfavors the hypothesis.

In the usual presentation of a scientific hypothesis, the normal procedure is to make the case in favor as strong as possible and leave criticism to others. Because the science of ball lightning is so little known in the wider scientific community, we have taken the unusual step of using what we know of that science to present both the pro and con cases. The much better-known physics of magnetic monopoles, although obviously incomplete observationally, has been treated with less detail.

The main intention of this paper is to gain the attention of those interested in magnetic monopoles to consider an admittedly unlikely, but non-zero-chance possibility: that an apparently unrelated but comparatively frequently-observed phenomenon—ball lightning—may be powered by the particle that has been sought for many decades: the magnetic monopole. As we show in an as-yet-unpublished article, the widespread use of security cameras is beginning to produce video recordings which may soon provide a relative abundance of objective data about ball lightning, including data that may favor the magnetic-monopole hypothesis. A deliberate effort directed toward obtaining more recordings and other data on ball lightning could yield critically important results, as evidence of ionizing radiation, magnetic susceptibility, or other characteristics of ball lightning that would favor the magnetic-monopole hypothesis would be highly significant. Theoretical studies of how thermal magnetic monopoles would interact with air could also lead to more specific predictions about how such an entity would behave.

If this paper convinces the reader that this hypothesis is worthy of further investigation, it will have served its purpose.